\documentclass[a4paper,fleqn]{cas-dc}

\usepackage[numbers]{natbib}
\usepackage{graphicx}
\usepackage{multirow}
\usepackage{amsmath,amssymb,amsfonts}
\usepackage{amsthm}
\usepackage{xcolor}
\usepackage[cal=rsfs]{mathalfa}
\usepackage{subcaption}

\def\tsc#1{\csdef{#1}{\textsc{\lowercase{#1}}\xspace}}
\tsc{WGM}
\tsc{QE}
\tsc{EP}
\tsc{PMS}
\tsc{BEC}
\tsc{DE}

\begin{document}
\let\WriteBookmarks\relax
\def\floatpagepagefraction{1}
\def\textpagefraction{.001}

\shorttitle{FEDSA: DLS in the Breast Cancer Screening}  

\shortauthors{Fernández, Gómez and Miranda}  

\title [mode = title]{Frequency Domain Analysis of Dynamic Light Scattering in the Breast Cancer Risk Screening: A Proof of Concept}
\tnotemark[1]

\tnotetext[1]{This document is the results of the research
   project funded by the \textit{Universidad Industrial de Santander}.}


\author[1]{Janeth Fernández-Pinto}
\credit{Conceptualization, Methodology, Investigation, Data Curation, Validation, Visualization, Writing - Original Draft, Writing - Review \& Editing}

\author[1,2]{Álvaro Gómez-Torrado}
\credit{Methodology, Resources, Writing - Original Draft, Writing - Review \& Editing}

\author[1]{David A. Miranda}[orcid=0000-0003-3130-3314]
\ead{dalemir@uis.edu.co}
\credit{Conceptualization, Methodology, Software, Data Processing, Data Curation, Investigation, Resources, Validation, Supervision, Visualization, Writing - Original Draft, Writing - Review \& Editing, Project administration}

\cormark[1]




\affiliation[1]{organization={CIMBIOS-CMN, Universidad Industrial de Santander},
            addressline={Cra 27 Cll 9}, 
            city={Bucaramanga},
            postcode={680002}, 
            state={Santander},
            country={Colombia}}

\affiliation[2]{organization={Empresa Social del Estado, Hospital Universitario de Santander},
            addressline={Cra. 33 No 28-126}, 
            city={Bucaramanga},
            postcode={680002}, 
            state={Santander},
            country={Colombia}}

\cortext[1]{Corresponding author}

\fntext[1]{}

\begin{abstract}
We present a proof of concept for screening breast cancer risk by detecting biochemical alterations in breast tissue using a noninvasive and low-cost technique that uses dynamic light scattering for field effect detection by spectral analysis (FEDSA). This technique consists of a light source that illuminates the tissue, and the backscattering light by the tissue is acquired by two detectors; next, the signal goes to the acquisition system, and then with the designed software the power spectra are calculated. The power spectra contain the frequency contribution related to the size of tissue compounds. These frequency contributions change with the biochemical alterations that are amplified by the field effect on tissue. This implies that the initial alterations in the breast are not local. To test FEDSA, two experiments were performed: the first was with Alumina particles grouped in average sizes of $60-300nm$, $100 - 400nm$ and polystyrene nanoparticles in suspension ($315nm$). The second was with 24 women, 17 of whom had normal tissue and 7 abnormal; abnormalities were previously detected by mammogram or ultrasound. Power spectra were obtained for all particles, in agreement with the particle size. In the case of normal and abnormal tissues, the power spectra of the tissues show differences in the shape of the spectra in the range of 1 to 160 kHz. ROC analysis suggests a possible good sensitivity (87.5\%) and specificity (68.1\%) in classifying breast tissue conditions. Statistical analysis with $p<0.05$ revealed significant differences in two quadrants of the breast, the upper inner right and the upper outer left, which were consistent with previous diagnoses by mammography and ultrasound. This proof of concept opens the possibility of implementing and improving FEDSA in the detection of early anomalies in the breast as a low-cost technique that does not use ionizing radiation.
\end{abstract}


\begin{highlights}
\item Proof of concept for screening breast cancer risk using dynamic light scattering.
\item Noninvasive, low-cost technique for field effect detection by spectral analysis.
\item Significant differences in power spectra between normal and abnormal breast tissues.
\item Potential good sensitivity and specificity in screening breast tissue conditions.
\item Opens the possibility for early detection of breast anomalies without ionizing radiation.
\end{highlights}


\maketitle


\section{Introduction}
\label{introduction}

Breast cancer is one of the most significant causes of cancer death in the world \cite{sung2021global}. Therefore, early detection plays a critical role in the preservation of human life, as has been demonstrated by applying the Papanicolaou test to the detection of cervical cancer, with a reduction in the risk of death \cite{buskwofie2020review}. To detect breast cancer, X-ray mammography images are the most widely used technique, but are not applicable to all risk populations \cite{aldhaeebi2020review}. Other technologies such as nuclear magnetic resonance, ultrasound, transillumination, and diffuse optical detect breast anomalies by imaging, but some of them require the maintenance of equipment and supplies at a high cost \cite{pal2020optical}.

Similarly to the Papanicolau test, which has been used successfully to detect cervical cancer \cite{kuhl2019abbreviated}, a technique applicable to the early detection of breast cancer is desirable. New promising methods for early detection of cancer include partial-wave spectroscopy (PWS) analysis of the spectra of partial waves propagating within a weakly disordered medium such as biological cells \cite{Vadim}. Electrical impedance tomography (EIT) is an imaging technique that maps the resistivity or the difference in resistivity of biological tissues from electrical signals \cite{mi13040496}. PanSeer is a non-invasive blood test based on circulating tumor DNA methylation \cite{chen2020non}. Photoacoustic computed tomography (3D-PACT) allows multipurpose imaging to reveal detailed angiographic information in biological tissues \cite{lin2021high}. iBreastExam is a handheld probe with an array of dynamic pressure sensors that measure tissue elasticity by making capacitive measurements on the breast surface and quantifying tissue stiffness variations \cite{mango2022ibreastexam}, and breastlight shines a frequency-specific safe and harmless red light through breast tissue, abnormalities are visibly identified as dark and shadowy spots \cite{labib2013evaluation}. 

Early cancer detection involves measuring alterations in the cellular microenvironment present before cancer appears as significant abnormal cell proliferation \cite{radosevich2013, curtius2018evolutionary}. Backman et al. have suggested that microenvironment alteration due to the presence of abnormal cells (cancer) is attributed to a field cancerization effect \cite{ROY2011, backman2011light}. The concept of a field effect in cancer originated in 1953 from histopathological observations by Slaughter et al. \cite{Slaughter}. In their work, they propose that the epidermoid carcinoma of the \lq\lq stratified oral squamous epithelium\rq\rq\, originates from a process of \textit{field cancerization}, in which an unknown carcinogenic agent has preconditioned an area of the epithelium (apparently normal). When an external agent, that is, a carcinogen, is applied, the time interval between application and cancer development varies depending on the intensity and duration of exposure to the agent; this exposure produces irreversible changes in individual cells or in groups of cells \cite{Slaughter}. Furthermore, the evolutionary perspective of \textit{field cancerization} by Curtius et al. associates this field with \lq\lq the replacement of the normal cell population by a cancer-primed cell population that may show no morphological change, is now recognized to underlie the development of many types of cancer\rq\rq \cite{curtius2018evolutionary}. This concept explains the multifocal character of cancer by proposing that various mechanisms within a patch or cell group can develop a carcinogenic \textit{field effect} at different tissue points. These foci can be microscopically or macroscopically related, leading to the production of multiple tumors in the tissue and extending along it until they invade further. The concept is also used to explain cancer recurrence, as surgery rarely goes beyond the limits of abnormal tissue, leading to changes in preconditioned tissue, or \textit{field cancerization}, which can turn into cancer again \cite{Slaughter, curtius2018evolutionary}.

\begin{figure}[t]
	\centering 
	\includegraphics[width=0.35\textwidth, angle=0]{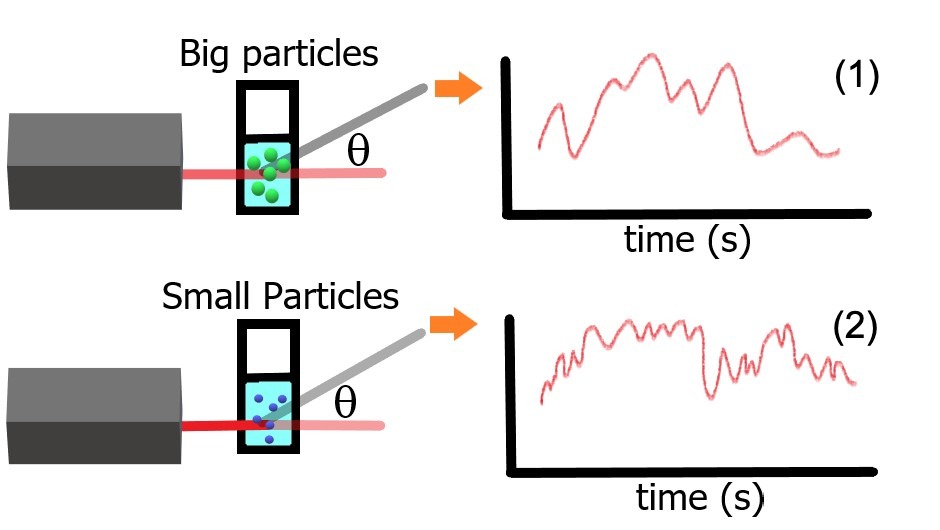}	
	\caption{Dynamic light scattering in the characterization of particles with different sizes, where the fluctuations in scattered intensity due to light scattering of particles describing a Brownian movement, differs from large particles, that exhibit low frequencies (1), and small particles, with high frequencies (2).} 
	\label{movement browniano.png}%
\end{figure}

Our interpretation of field cancerization was proposed in 2016 as an amplification of biochemical abnormalities in tissues \cite{Fernandez2016}. This interpretation considers that all tissue alteration can be explained as the product of biochemical changes that we call biochemical abnormalities, and that normal tissue amplifies, through a complex signaling process, abnormalities initiated by abnormal cells (i.e. the transformed cells that become, after time, into cancer). With this interpretation, the assessment of breast cancer risk by image processing could be explained as a consequence of nonlocal changes induced by the field cancerization effect (or simply the \textit{field effect}) in affected tissue \cite{Miranda-Pertuz2020} and can be used to explain the association observed between mammographic parenchymal patterns and the risk of breast cancer and the risk assessment of breast cancer \cite{saslow2007, giger2013, onega2014, vilmun2020, Hernndez2023}.

Here, we present \textit{field effect} detection by spectral analysis (FEDSA) as an experimental technique to indirectly measure \textit{field cancerization} and its application to the early detection of breast cancer.  The proposed method was inspired by the interpretation of the field cancerization effect as an amplification of cellular abnormalities, in which the amplification effect can be used for the early detection of cancer by measuring the change in cell dynamics proposed by Fernandez-Pinto et al. \cite{Fernandez2016}.
\subsection{Field Effect Detection by Spectral Analysis (FEDSA)}

\begin{figure}[t]
	\centering 
	\includegraphics[width=0.4
 \textwidth, angle=0]{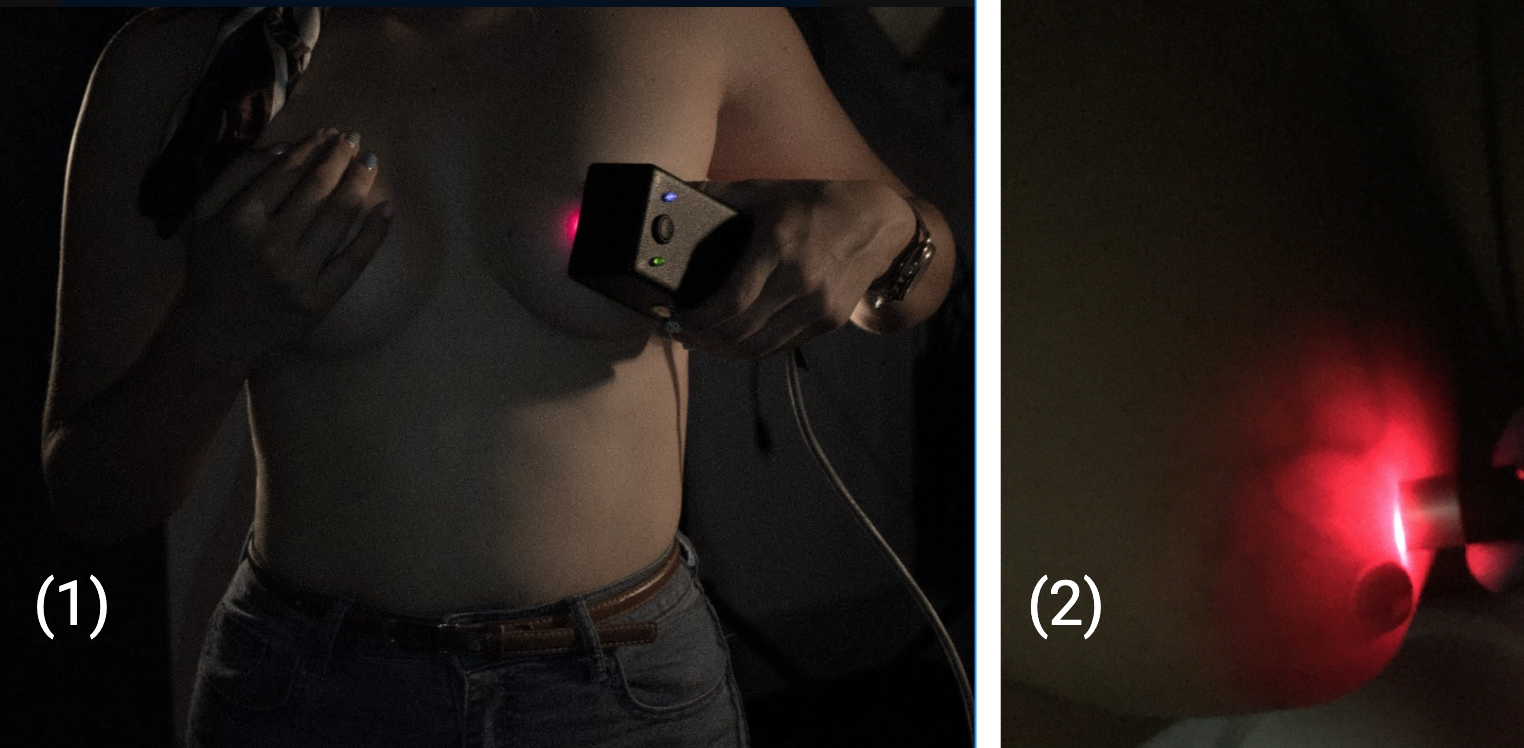}	
	\caption{ Emitter and detectors (black box) of the FEDSA system placed on the breast (1). The backscattered light contains signals from nearby cells (2), but due to field effect amplification, it also provides information on the state of internal breast alterations.} 
	\label{capas del tejido mamario.jpg}%
\end{figure}

The development of FEDSA technology was inspired by accessing the molecular size distribution by analyzing a signal from particles describing a Brownian movement, such as molecules in a tissue; see Figure~\ref{movement browniano.png}. The signal is obtained from the continuous NIR light scattered by the breast \cite{MirandaFernandez}. In this investigation, it has been hypothesized that the changes generated by abnormalities located within breast tissue are not only found around them but spread throughout the tissue, i.e. the carcinogenic field effect is not local. This implies that the appearance of some abnormalities will modify all cell signaling within the tissue \cite{Fernandez2016}. These changes in cell signaling have already been verified in breast cancer \cite{mccleskey}, but there are still no optical techniques that can directly measure the emergence or suppression of all proteins involved in the process. A possible first step is to measure the amplification of abnormalities, through patients with clinically verified alterations; this way, when comparing the cellular dynamics of mammary tissues of a healthy and abnormal breast, it is expected to observe that dynamic breast tissue has changed.

To study the dynamics of molecules in tissue, the information collected is mainly recovered from NIR light backscattered by breast tissue at the time. This light is in its path a layer of dead cells called the stratum corneum, which produces the reflection of the beam between $5 - 7\%$, these are located within the epidermis, which is a layer of $0.027 - 0.15mm$ \cite{anderson1981optics,meglinsky2001modelling,doi2003spectral,bara}. The epidermis has absorption properties dominated by melanin, a natural chromophore produced by melanocyte cells, and has scattering properties. These two skin layers are characterized by forward scattering of light. Subsequently, the dermis $0.6 - 3mm$ is found, which is composed of irregular connective tissue, nerves, and blood vessels \cite{anderson1981optics,meglinsky2001modelling,doi2003spectral,bara}. Then there is the superficial fascia, which is made up of supraglandular connective tissue, Cooper ligaments, and blood vessels, and could finally reach the interglandular adipose tissue. These last two layers do not have specific size values, as this varies from body to body. In some women, NIR light can only reach the superficial fascia and in others it can reach the adipose tissue; this is due to the fact that NIR light penetrates to a depth of centimeters, see Figure~\ref{capas del tejido mamario.jpg}, and the size of the mammary gland from the nipple to the pectoral in a young woman is on average $21 cm$ \cite{rehnke}.

\subsection{Brownian Movement and Light Scattering}

\begin{figure}[t]
    \centering
    \includegraphics[width=0.45\textwidth, angle=0]{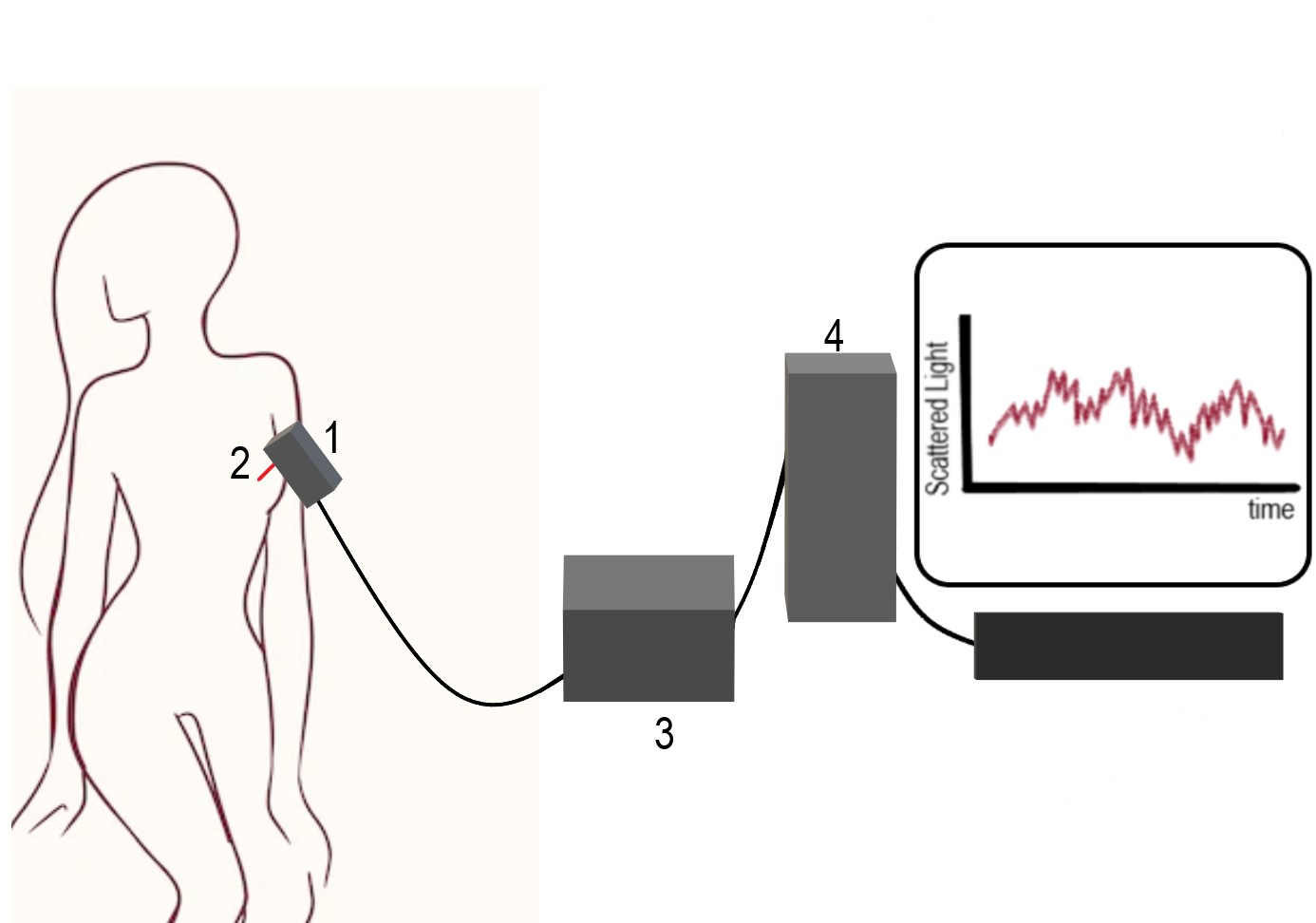}
    \caption{FEDSA system utilized for characterizing human breast tissue. The experimental setup includes (1) an illumination and detection system, (2) the breast tissue being measured, (3) a data acquisition system, and (4) a data processing system.}
    \label{FEDSACONFIGURATION.png}
  \end{figure}

Based on the hypothesis that biochemical abnormalities caused by abnormal cells and amplified by normal ones can indicate the presence of the field cancerization effect, as previously proposed \cite{Fernandez2016}, we suggest that the size distribution of biological molecules in normal and abnormal tissues may differ and could be correlated with cancer risk. In this context, evaluating the size (or mass) distribution of biological molecules in tissue can serve as a method to detect abnormalities related to the field cancerization effect. To test this hypothesis, we developed the FEDSA system, drawing inspiration from the Brownian motion of molecules, where smaller (lighter) and larger (heavier) molecules exhibit distinct velocity distributions. We employ spectral analysis of light scattering to ascertain the distribution of molecules within tissue on the basis of variations in their sizes (and mass).

\section{Materials and methods}

\subsection{Experimental Design of FEDSA Technique}

The FEDSA device was designed to measure the backscattered light from breast tissue, as shown in Figure~\ref{FEDSACONFIGURATION.png}. The detected signal is viewed on a graphical interface as a voltage-versus-time signal. The system has a light emitter and two detectors; these are inside a black box with two buttons that turn on the emitter and another that activates the detectors; for measurements in breast tissue, a dark room is required for robust measurements. The signal obtained by each detector and its difference go to a data acquisition system. 

\subsubsection{Materials}
The illumination system has an emisor diode LED with a central wavelength at 657 nm, 0.5 mW output power and
operating at 35.5$^{\circ}$ 
average temperature. Two detectors ODA-5WB-100K with an active area of 5mm$^{2}$, with a $\pm 9 V$ power supply, a 500 kHz bandwidth and a response of 40 V / mW (for the implementation of LEDs) were used for the detection of scattered light. The emitter and detectors (black box) of the FEDSA system are placed on the breast; see Figure~\ref{capas del tejido mamario.jpg}. Data acquisition was controlled by a compact cRio 9067 system with
the 9223 module coupled inside the chassis. Module 9223 has four channels to record input signals with a rate of acquisition $1MS/s$, which is distributed by the three channels used (approximately $333\, kHz$ by channel).

\subsubsection{Hardware}
The materials were placed in a configuration as shown in Figure~\ref{FEDSACONFIGURATION.png}. The illumination and detection system (1) was placed inside a black box, which has three holes equidistantly separated; in the middle is the emitter and on each side is a detector. An amplification with gain $G = 11$ was implemented with an operational amplifier. In the interaction of light with tissue (2), three signals are measured. Two correspond to the detectors, while the third is the difference between these two, with the latter
being amplified. Data are acquired by the compact Rio 9067 system, using the 9223 module (3). This module acquires each signal (one per channel) at a sampling frequency of $333\, kHz$, then, according to the Nyquist frequency, the maximum acquisition frequency will be approximately $166\, kHz$. All data are stored on the computer where they are processed (4).

\subsubsection{Software}
The interaction between the illumination detection system and the data acquisition system was carried out by implementing
Labview software. Expert algorithms were implemented with Python. \footnote{Experimental data and a simple Jupyter Notebook that show the data processing is available at \cite{DavidAM}.}.
\subsection{FEDSA Proof of Concept }

\subsubsection{First Experiment: Particles in suspension}

To explore the correlation between backscattered light and particle size using the FEDSA system, an experiment with suspended particles of varying sizes was conducted. The particles utilized were of two types: alumina and polystyrene.

For the alumina particles, samples were obtained from different batches. Their sizes were determined using the Zeta-sizer Nano ZS (Z-sizer) from Malvern Instruments, revealing size ranges of $60-300\, nm$ and $100-400\,nm$. Polystyrene particles, known to be $315.9\,nm$ in size, were used as calibration standards for the Z-sizer.

To prepare the alumina suspensions, $3\,ml$ of distilled water was mixed with $30\,mg$ of alumina for each particle size range. These suspensions were placed in a quartz cuvette located $2\,cm$ from the light emitter. Backscatter detection was performed at a $176^{\circ}$ angle relative to the incident beam, with the sample oriented at a $4^{\circ}$ angle from the illumination and detection system.

We performed FEDSA measurements to acquire the power spectra of light scattered by the suspended particles. Each sample was subjected to measurements in triplicate to ensure reliability and consistency. The duration for each measurement was 10 seconds. The same procedure was applied to polystyrene particles for calibration purposes.

In addition, the alumina particles were examined using a Scanning Electron Microscope (SEM). The specific SEM model used in this study was the Quanta 650 Field Emission Gun (FEG).


\subsubsection{Second experiment: Breast tissue}

\begin{table*}[th]
\centering
\caption{Summary of diagnoses for the seven women with breast abnormalities, with patient numbers in the first column and detailed diagnoses in the second.}
\label{table:patients}
\begin{tabular}{>{\centering\arraybackslash}m{1.5cm} m{0.8\linewidth}}
\hline
Number & Diagnosis \\ \hline\hline
1 & Mammography and ultrasound indicate a nodular image in the upper outer quadrant of the right breast, corresponding to the grouped microcysts. Nodular images in the right quadrant with dimensions $7 \times 5 \times 4\ mm^3$, $5 \times 2 \times 3\ mm^3$, and $3 \times 5 \times 4\ mm^3$. \\ \hline

2 & Breast ultrasound reveals solid and characteristic benign nodules in the upper and lower internal quadrants of the left breast with approximate dimensions $12 \times 7 \times 6\ mm^3$. \\ \hline

3 & Mammography identifies masses in the upper quadrants of both breasts. The biopsy results show fibroadenoma in the right and left glands, measuring $3.0 \times 3.0 \times 2.0\ mm^3$ and $1.7 \times 1.0 \times 1.5\ mm^3$, respectively. \\ \hline

4 & Mammogram displays cysts in the upper right outer quadrant, measuring $6.6 \times 9.0 \times 3.6\ mm^3$ and $6.0 \times 7.0 \times 4.0\ mm^3$ at $4.5\ cm$ from the nipple. \\ \hline

5 & Mammography shows micro and macrocalcifications in the lower internal quadrant of the left breast. \\ \hline

6 & Mammography detects cysts in specific quadrants of the breast: UOQ (right) with $17.1 \times 5.9 \times 14.2\ mm^3$ at $2\ cm$ from the areola and UOQ (left) with $10.9 \times 5 \times 11.5\ mm^3$ at $2\ cm$ from the areola, UIQ (left) with $0.7 \times 5.6 \times 11.8\ mm^3$, LIQ (left) with $8.9 \times 4.4 \times 12.9\ mm^3$ at $4.5\ cm$ from the areola. The patient has single lumen gel-filled breast implants prefilled with silicone gel. \\ \hline

7 & During the measurement process, the patient reported abnormalities in breast tissue, but it was impossible to contact her after taking measurements. \\ \hline

\hline
\end{tabular}
\end{table*}

\begin{figure}[th]
    \centering
    \includegraphics[width=0.45\textwidth, angle=0]{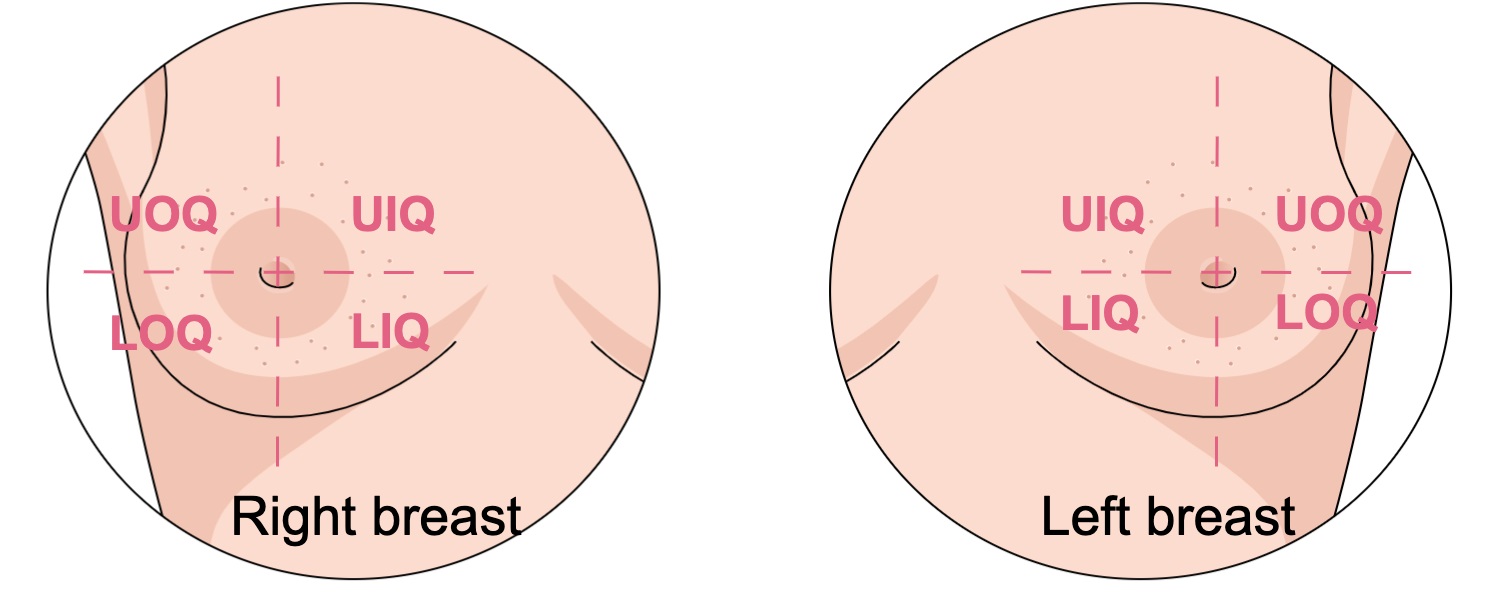}
    \caption{For the measurement process, the breast is divided into eight quadrants, four in each breast: Upper-Outer (UOQ), Upper-Inner (UIQ), Lower-Outer (LOQ), and Lower-Inner (LIQ).}
    \vspace{\baselineskip}
    \label{Breast-quadrants-definition.png}
\end{figure}

To evaluate the potential of FEDSA to differentiate normal from abnormal breast tissue, 24 women with an average age of $41 \pm 12$ years participated in the study. Of these, only seven women had previously been diagnosed with breast abnormalities. Six of these women had palpable masses and were aware of their diagnoses, while the seventh participant did not provide clinical reports, as described in Table~\ref{table:patients}. None of the participants had undergone chemotherapy or radiation treatment. Data were recorded according to the definition of breast quadrants: Upper-Outer (UOQ), Upper-Inner (UIQ), Lower-Outer (LOQ), and Lower-Inner (LIQ), as illustrated in Figure~\ref{Breast-quadrants-definition.png}. 

This pilot study received approval from the ethics committee named \textit{Comité de Ética en Investigación Científica} (CEINCI) at the \textit{Universidad Industrial de Santander} (UIS), Bucaramanga, Colombia. Informed consent was thoroughly explained and the measurement protocol was only started once the women provided their explicit acceptance and assignment. Participants had the option to withdraw at any time if they so desired. The research team adhered to the principles of the Helsinki Declaration to ensure the protection of the women's life, health, dignity, integrity, right to self-determination, privacy, and confidentiality.

All participants rested for 15 minutes before measurements, during which informed consent was explained and personal data was collected. The measurement process was conducted as follows:

\begin{itemize}
    \item First: A health professional divided the breast of each woman into quadrants as shown in Figure~\ref{Breast-quadrants-definition.png}.
    \item Second: The illumination and detection device was placed on the upper external quadrant, as illustrated by the self-positioning of the device in Figure~\ref{capas del tejido mamario.jpg}.
    \item Third: The tissue was illuminated for 10 seconds. The duration of illumination could vary depending on the desired level of detail.
    \item Fourth: The detection system captured backscattered light from the tissue.
    \item Fifth: The signal was stored and transferred to a computer.
    \item Sixth: Steps 2 to 5 were repeated for all quadrants of both breasts.
\end{itemize}

\subsection{Data Analysis}

In line with the hypothesis of abnormality amplification due to the field effect \cite{Fernandez2016}, the backscattered light signal measured by FEDSA is believed to contain information related to the size of molecular components within the intracellular and extracellular matrices of breast tissue. This concept suggests that larger molecular components will exhibit slower movement, and thus lower frequencies, in contrast to smaller components, which move more rapidly at higher frequencies. Although the total acquisition time of the signal affects the measurable frequency ranges, the motion frequencies of the molecules are determined by the size and velocity of the particles in suspension, not by the measurement duration.

To extract the relevant information, the power spectrum was computed. This process begins with a discrete signal analysis, followed by the application of white noise reduction techniques. The implemented approach is similar to the Welch method, which estimates the power spectrum of a signal at various frequencies by transforming the signal from the time domain to the frequency domain \cite{Welch}. Statistical analysis was also performed. The detailed methodology is provided in the following sections.

\subsubsection{Discrete signal analysis}

\begin{figure*}[t]
\centering
\includegraphics[width=0.96\textwidth, angle=0]{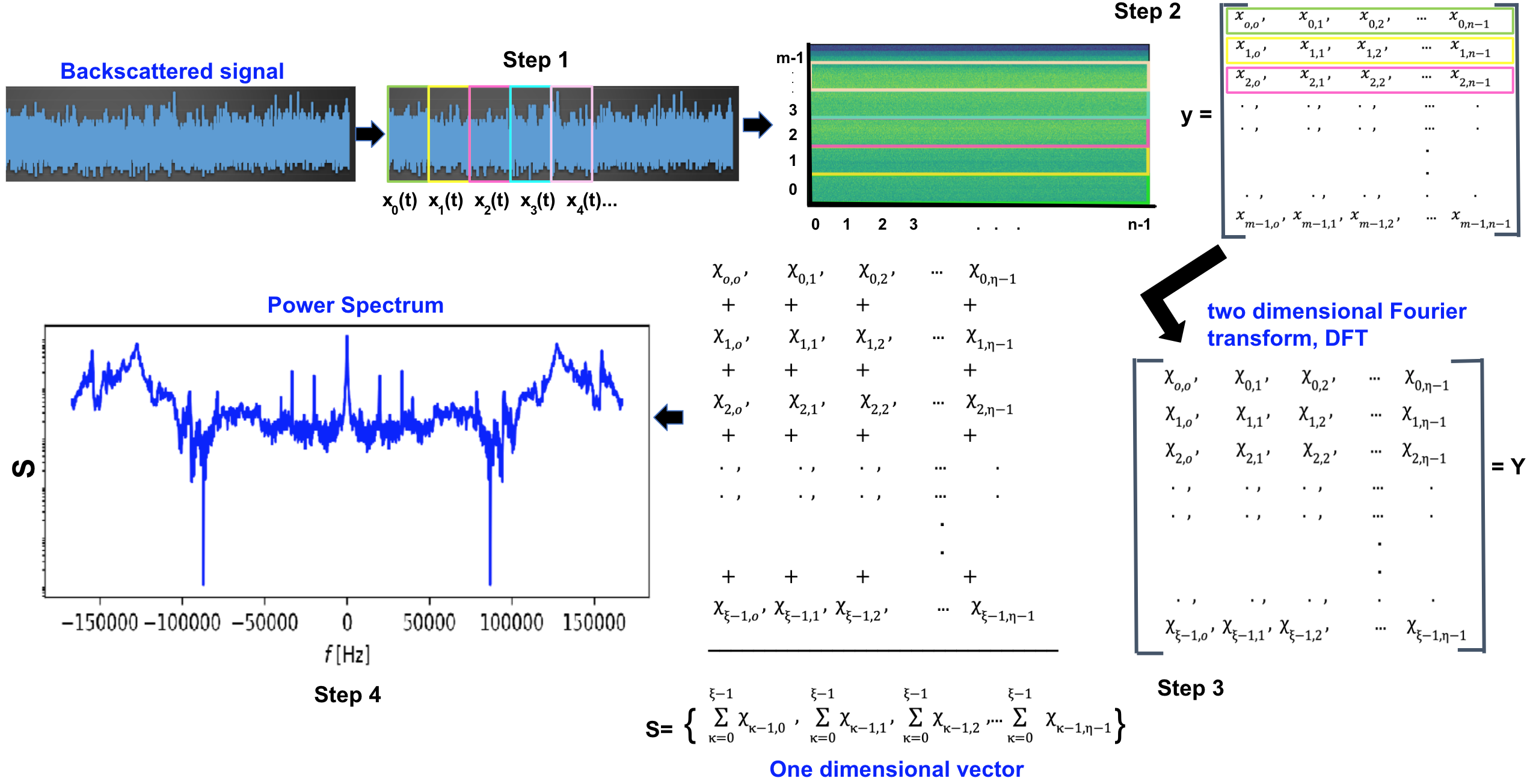}

\caption{Procedure for calculating the Power Spectrum: In Step 1, the voltage signal $x(t)$ is divided into $m$ windows, and the digital version of each $x_i(t)$ signal contains an equal number $n$ of data points. Step 2 involves vertically arranging these $m$ windows to form a matrix, represented as $y=\{x_{i,j}\}$. Step 3 applies a discrete Fourier transform to this matrix, producing a new matrix $Y=\{\chi_{k,l}\}$. To reduce noise, the rows of $Y$ are summed. Finally, Step 4 computes the power spectrum $S$.}
\label{tratamientodedatos.PNG}

\end{figure*}

The biological signal acquired over time, $x'(t)$, was digitally sampled as $x'[k]$ and normalized to $x[k]$ by Eq.~(\ref{eq:x-normalized}), where $k = 0, 1, \cdots, m \times n$ and $\langle x' \rangle$ is the mean of $x'$.

\begin{equation}
    x[k] = \frac{ x'[k] - \langle x' \rangle } { \max\left\{ \left| x'[k] - \langle x' \rangle \right| \right\} }
    \label{eq:x-normalized}
\end{equation}

This normalized signal, $x[k]$, was segmented into $m$ equal time intervals, each containing $n$ samples, as illustrated in Step 1 of Figure~\ref{tratamientodedatos.PNG}, where the digitized $x_i(t)$ signal, represented by colored rectangles in Step 1, has samples of the form $x_{i,j}$. Although Step 1 of Figure~\ref{tratamientodedatos.PNG} aligns with methodologies used in other studies, such as the analysis of data signals using a window function \cite{Lalith}, our approach differs in that we treated the representation $y[i, j]$ as an image, where $i = 0, 1, \ldots, m-1$ and $j = 0, 1, \ldots, n-1$. The matrix $y$, consisting of elements $y[i, j]$, is defined by Eq.~(\ref{eq:y-matrix}), where $x[ni + j]$ is one of the $m \times n$ digitized samples; i.e., for a given window $i$, the sample is identified as $x_{i,j}$ in Step 2 of Figure~\ref{tratamientodedatos.PNG}. Here, the rectangle function ($rect[k/n]$), as described in Eq.~(\ref{eq:rect}), performs the division of the signal $x[k]$ into $m$ digital signals, corresponding to the $m$ windows, each with $n$ samples.

\begin{equation}
    y[i,j]=x[ni + j] rect\left[\frac{ni + j - n/2}{n}\right]
    \label{eq:y-matrix}
\end{equation}

\begin{equation}
    rect \left[\frac{k}{n} \right] = 
    \left\{ \begin{array}{ll}
    1 & \mbox{if\, $|k| \leq n/2$}\\
    0 & \mbox{if\, $|k| > n/2$}.\end{array} \right. 
    \label{eq:rect}
\end{equation}

Once the matrix $y$ is obtained, the discrete Fourier transformation is applied and its square magnitude is stored in a new matrix $Y$, as detailed in Eq.~(\ref{eq:Y-matrix}) and illustrated in Step 3 of Figure~\ref{tratamientodedatos.PNG}. Furthermore, to minimize noise, a new signal $S$ was constructed, its elements being the sum of the rows in $Y$; this is outlined in Eq.~(\ref{eq:S-matrix}). The signal $S$ represents a low noise power spectrum, analogous to that obtained using the Welch method, and functions as a frequency-dependent variable, as shown in Step 4. The power spectrum $S$ is displayed on a logarithmic scale, as shown in Step 4 of Figure~\ref{tratamientodedatos.PNG}.

\begin{equation}
Y[\kappa, l] = \chi_{\kappa, l} = \{|DFT(y)|^2\}_{\kappa,l}
\label{eq:Y-matrix}
\end{equation}

\begin{equation}
S[l] = \sum_{\kappa=0}^{\xi-1}Y[\kappa, l]
\label{eq:S-matrix}
\end{equation}

\subsubsection{Power spectrum bands}

In this study, a spectral band $b_{i}^{(k)}$ from frequency $f_i$ to frequency $f_i + \Delta f_i$, for experiment $k$, is defined by Eq.~(\ref{eq:bands}), where $1/\gamma$ represents a normalization factor that ensures that the sum of all bands for the experiment $k$ -th is equal to one, that is, $\sum\limits_i b_{i}^{(k)} = 1$.

\begin{equation}
    b_{i}^{(k)} = \frac{1}{\gamma} \int\limits_{f_i}^{f_i + \Delta f_i} \ln \{ S(f) \} \, df
\label{eq:bands}
\end{equation}

\begin{equation}
    \frac{1}{\gamma} = \int\limits_{-\infty}^{\infty} \ln \{ S(f) \} \, df
\end{equation}

The trapezoidal rule was used to numerically calculate the integrals that define each band $b_{i}^{(k)}$ using the discrete power spectrum $S[l]$ defined by Eq.~(\ref{eq:S-matrix}), within the frequency interval between $f_i$ and $f_i + \Delta f_i$ for experiment $k$.

\subsubsection{FEDSA Spectral parameters and field cancerization effect}

The FEDSA system stored the information obtained from the two sensors (D1 and D2) and their difference (Dif) as digital signals. Three data acquisition channels were used, one for each signal (D1, D2, and Dif). For each participant, data were recorded from the eight quadrants of breast tissue (Figure~\ref{Breast-quadrants-definition.png}) for 10 seconds per quadrant and per channel. A total of 576 measurements were obtained as follows: for each of the three channels (D1, D2, and Dif), in one of the quadrants of breast tissue of a woman, the procedure described in Figure~\ref{tratamientodedatos.PNG} was applied. For each of the 24 participants included in the study, we measured three signals in each of the eight quadrants (four on the left and four on the right). Taking into account four windows for data analysis (512, 1024, 2048, and 4096), four power spectra were obtained for each experimental data set. This process was repeated for the 24 women, resulting in a total of 2304 power spectra.

The normalized $i$-th area under the power spectrum curve from frequencies between $f_i$ and $f_i + \Delta_i$, also known as the \textit{band}, for the $k$-th experiment was calculated using Eq.~(\ref{eq:bands}), which defined six $b_i^{(k)}$ \textit{bands} within specific frequency ranges defined in Table~\ref{tab:frequency-ranges}. All data were organized for subsequent statistical analysis to determine whether there were significant differences between the measurements obtained for the group of women classified as normal and those classified as abnormal.

\begin{table}[h]
\centering
\caption{Frequency ranges for the six spectral bands.}
\label{tab:frequency-ranges}
\begin{tabular}{ccc}
\hline 
Band $b_i^{(k)}$ & $f_i$ $[kHz]$ & $f_i + \Delta f_i$ $[kHz]$ \\
\hline \hline
$b_1^{(k)}$ & 0 & 1 \\
$b_2^{(k)}$ & 1 & 10 \\
$b_3^{(k)}$ & 10 & 50 \\
$b_4^{(k)}$ & 50 & 100 \\
$b_5^{(k)}$ & 100 & 150 \\
$b_6^{(k)}$ & 150 & 160 \\
\hline
\end{tabular}
\end{table}


\subsubsection{Multivariate Analysis and Logistic Regression Modeling}

Data obtained from the FEDSA experiment, specifically \textit{ bands} $b_1^{(k)}$ to $b_6^{(k)}$, were classified into two groups: normal (0) and abnormal (1). These classifications were based on the signals from each detector (D1 and D2) and their difference (Dif), the window sizes (512, 1024, 2048, 4096) and the quadrants (UOQ, UIQ, LOQ, LIQ for both the right and left sides). A multivariate analysis was performed to identify potential differences between these groups, focusing on a dichotomous dependent variable that represents the condition of breast tissue as normal (0) or abnormal (1). This analysis aimed to evaluate the significant differences between the groups.

The normality of the groups was assessed using the Shapiro-Wilk test and homoscedasticity was verified with the Levene test. If the data met both normality and homoscedasticity criteria, a t-student test was applied. If normality was not met, an independent sample Mann-Whitney U test, a nonparametric statistical test, was used \cite{mcknight2010mann,mala2021power}.

Subsequent to the multivariate analysis, a logistic regression model was used to differentiate between normal and abnormal breast tissue conditions. To avoid linear dependence due to the condition $\sum\limits_i b_{i}^{(k)} = 1$ for the six \textit{bands}, a linear transformation was applied as per Eq.~(\ref{eq:LinearTransformationPCA}), derived from a principal component analysis (PCA). This transformation reduced the dimensionality of the data set and produced five linear independent principal components $\mathfrak{b}_j^{(k)}$. The linear independence of the components was evaluated using the variation inflation factor (VIF).

\begin{equation}
    \label{eq:LinearTransformationPCA}
    \mathfrak{b}_j^{(k)} = \sum_{i=1}^6 T_{ji} b_{i}^{(k)}
\end{equation}

The performance of the logistic regression model, as described by Eq.~(\ref{eq:log_reg_model}) and incorporating the model parameters $\beta_j$, was assessed by a Receiver Operating Characteristic (ROC) analysis. This analysis aimed to determine the efficacy of the model in distinguishing between normal and abnormal breast tissue conditions. Sensitivity and specificity were used to evaluate model performance, with the optimal cutoff point identified using the Youden J statistic.

\begin{equation}
    \mathfrak{p}_k = \left[ 1 + e^{\beta_0 + \sum\limits_{j=1}^5 \beta_j \mathfrak{b} _j^{(k)}} \right]^{-1}
    \label{eq:log_reg_model}
\end{equation}

\begin{figure}[t]
    \centering
    \includegraphics[width=0.45\textwidth, angle=0]{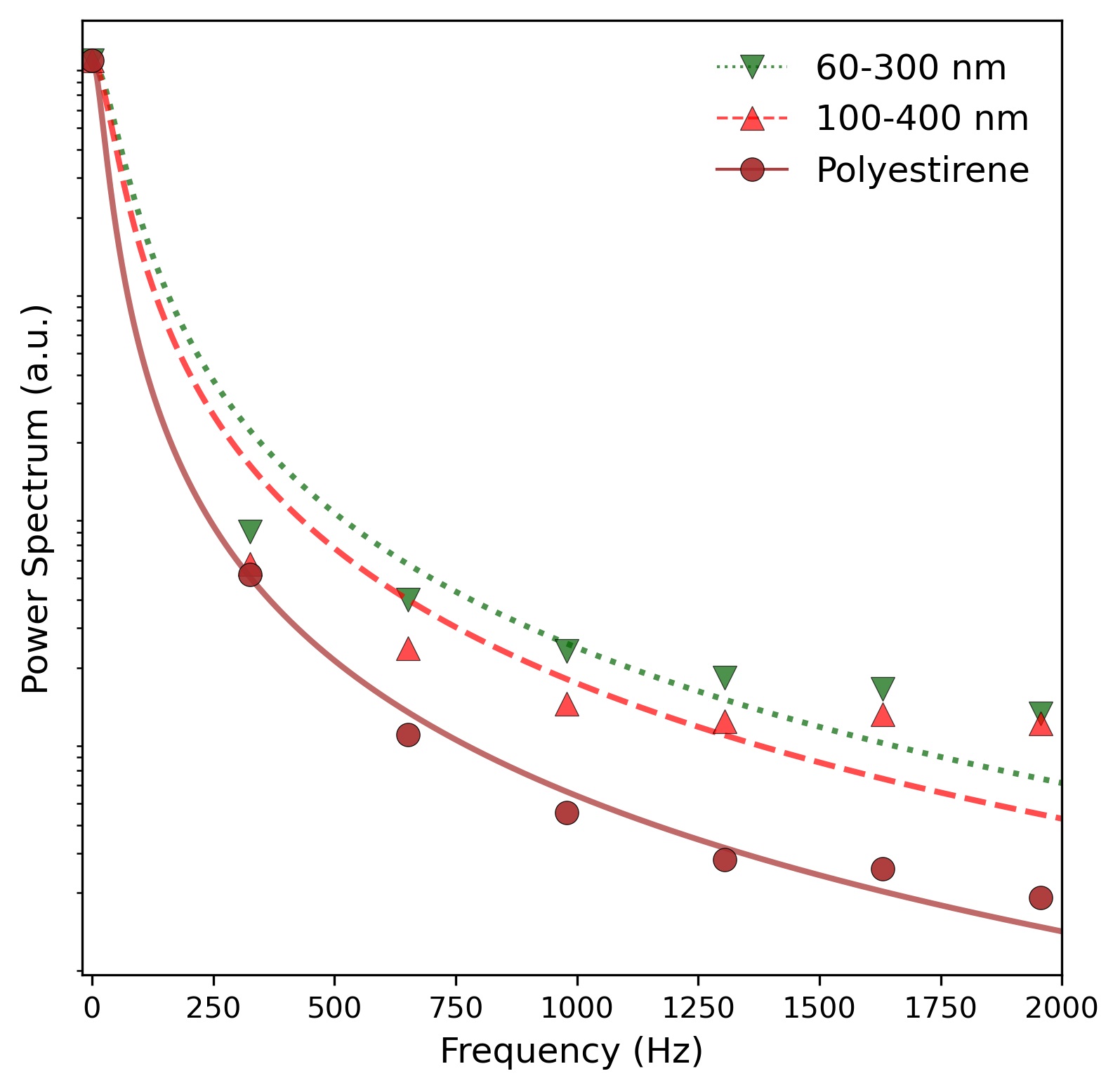}
    \caption{Power spectrum as a function of frequency, obtained with FEDSA, for particles of different sizes suspended in solution. The dotted line ($\cdots$) represents the fitting to Eq.~(\ref{eq:PowerSpectrum_BrownianMovement}) of data for a mixture of alumina particles with sizes in the range $60-300\,nm$; the dashed line ($--$) for the range $100-400\,nm$, and the solid line for polystyrene particles with an average size of $315.9\,nm$.}

    \label{SPP.png}
\end{figure}

\section{Results}

\subsection{FEDSA Applied to Particles in Suspension}

\begin{figure}[t]
    \centering
  \includegraphics[width=0.475\textwidth, angle=0]{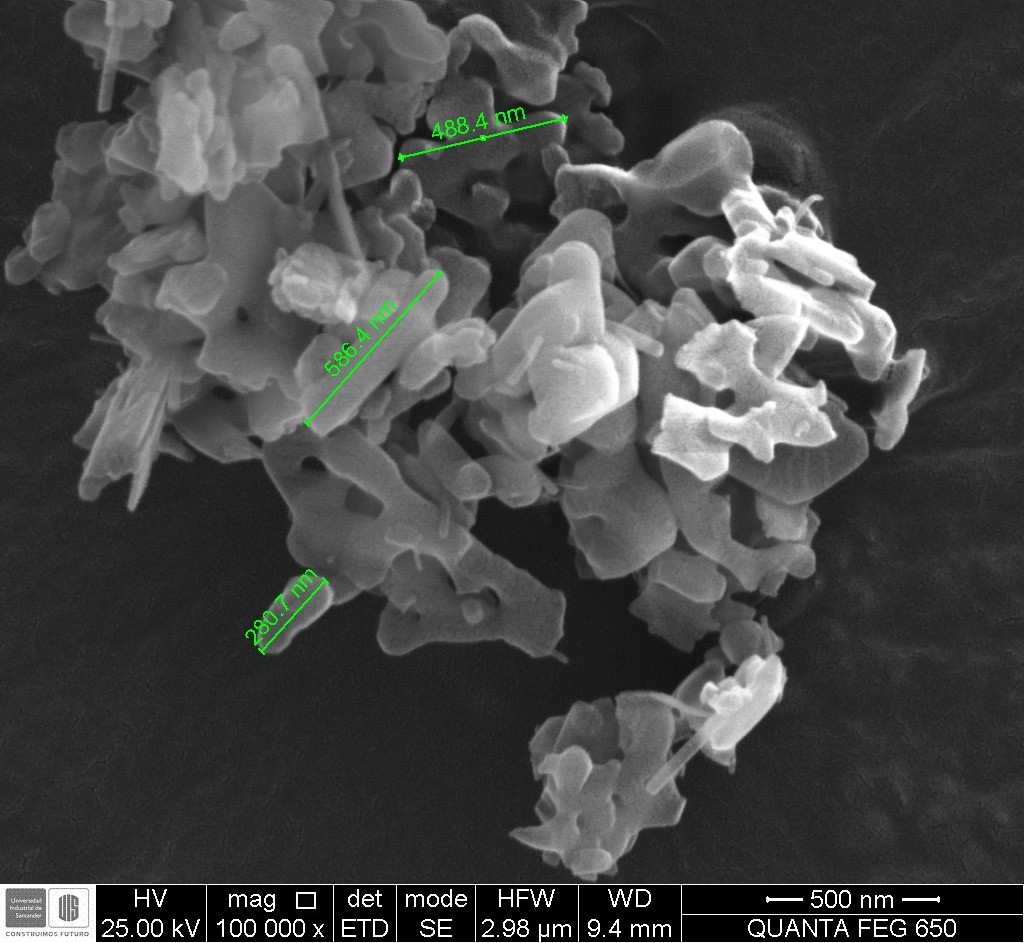}
    \caption{Scanning electron microscopy (SEM) images used to determine the sizes of the Alumina particles. The image corresponds to the nanoparticles in powder form before being sonicated to separate them for the experiments.}
    \label{SSPp.jpg}
\end{figure}

The power spectrum obtained with FEDSA for particles suspended in a solution is shown in Figure~\ref{SPP.png}, and the sizes of some alumina particles, as observed by scanning electron microscopy (SEM), are presented in Figure~\ref{SSPp.jpg}. The FEDSA data were processed according to the algorithm described in the Materials and Methods section, resulting in power spectra $S[l]$. For particles undergoing Brownian motion in suspension, the experimental data in the frequency range from $0\,\text{Hz}$ to $2000\,\text{Hz}$ were normalized to the lower frequency $\omega_0$ and fitted (as shown by the lines in Figure~\ref{SPP.png}) to the power spectrum of classical particles that describe Brownian motion in the long-term approximation, as given by Eq.~(\ref{eq:PowerSpectrum_BrownianMovement}) \cite{Kubo1991}. Here, $\omega = 2\pi l f_s$ represents the angular frequency, $f_s$ the sampling frequency, $S(\omega) = S[l]$, $\omega_0$ the characteristic frequency as per Eq.~(\ref{eq:characteristic_freq}), and $\delta$ is a model parameter. 

\begin{equation}
    \label{eq:PowerSpectrum_BrownianMovement}
    \frac{S(\omega)}{S(\omega_0)} = \frac{e^\delta}{\pi} \frac{\omega_0}{\omega^2 + \omega_0^2}
\end{equation}

\begin{equation}
    \label{eq:characteristic_freq}
    \omega_0 = \frac{a}{r}
\end{equation}

The parameter values obtained by fitting the experimental data to Eq.~(\ref{eq:PowerSpectrum_BrownianMovement}) are listed in Table~\ref{table:characteristic_frequency}. The value of $a \approx 46100\,\text{nm}\,\text{rad/s}$ was determined by substituting the expected size of the calibration polystyrene particles ($315.9\,\text{nm}$) and the characteristic frequency $\omega_0$ from the fitting into Eq.~(\ref{eq:characteristic_freq}). Theoretically, $a$ is expressed as $8\pi k_B T \sin^2(\theta / 2) / (3\lambda^2\eta)$, where $\lambda$ is the wavelength in suspended medium, $\eta$ the viscosity, $\theta$ the scatter angle, $r$ the radius of the particles, $k_B$ the Boltzmann constant, and $T$ the temperature.

\begin{table}[t]
\caption{Parameters of the power spectrum shown in Figure~\ref{SPP.png} for particles in suspension, described by the Eq.~(\ref{eq:PowerSpectrum_BrownianMovement}), which characterizes particles undergoing Brownian movement. The table also presents the predicted particle sizes (Size $[nm]$) obtained using Eq.~(\ref{eq:characteristic_freq}) assuming the polystyrene particles as calibration ones.}
\label{table:characteristic_frequency}
\centering
\begin{tabular}{cccc}
\hline
Particles & $\omega_0$ $[rad/s]$ & $\delta$ & Size $[nm]$\\
\hline\hline
Alumina $(60-300)$ nm & 316& 6.87 & 146 \\
Alumina $(100-400)$ nm & 262& 6.70 & 177\\
Polystyrene $\sim 315.9$ nm & 146& 6.13 & 316 \\
\hline
\end{tabular}
\end{table}

In our analysis, we assumed that $a$ remains constant regardless of the type of sample. Therefore, we derived the parameter $a$ from the known sizes of the polystyrene calibration particles, which allows us to estimate the sizes of the alumina particles, as indicated in Table~\ref{table:characteristic_frequency}. As expected, the fit to the theoretical model was accurate for the calibration particles (polystyrene), which closely resemble the sizes predicted by the theoretical model, as seen in Figure~\ref{SPP.png}. Furthermore, comparing the sizes obtained using the parameter $a$ (column labeled \lq\lq Size $[nm]$\rq\rq) with those reported by the Zeta-sizer Nano ZS system (column labeled \lq\lq Particles\rq\rq, in Table~\ref{table:characteristic_frequency}), we found a close approximation using FEDSA. This result validates our hypothesis that FEDSA measurements can provide information on particle sizes\footnote{Experimental data and a simple Jupyter Notebook demonstrating the data processing are available at \cite{DavidAM}.}.

\subsection{FEDSA Applied on Breast Tissue}

\begin{figure*}[t]
    \centering
    \begin{subfigure}[b]{0.45\textwidth}
        \includegraphics[width=\textwidth, angle=0]{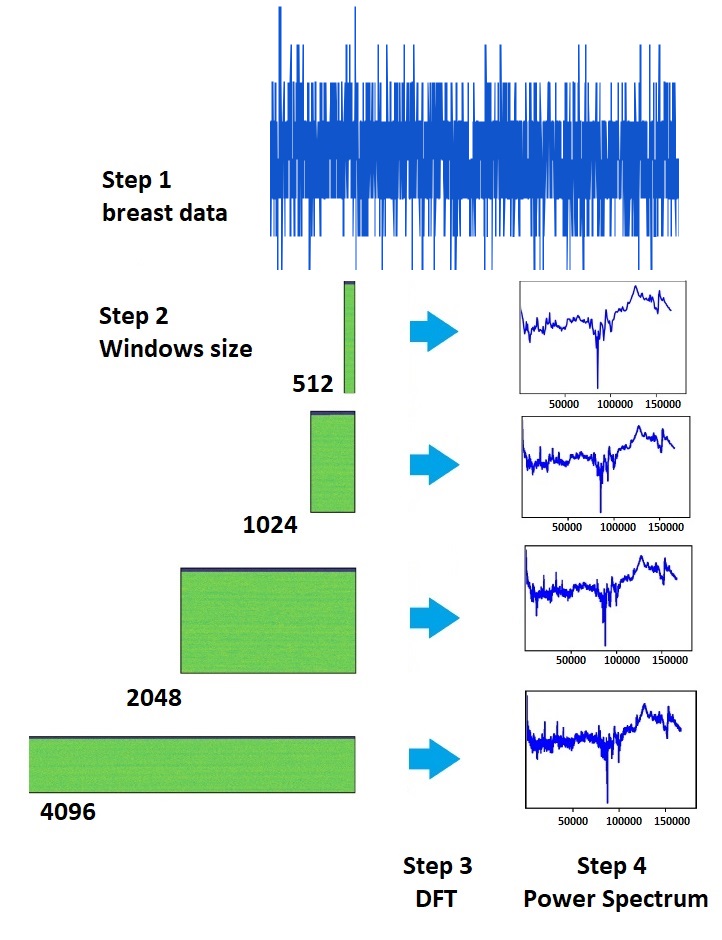}
        \caption{}
        \label{fig:abnormal_breast}
    \end{subfigure}
    \hfill 
    \begin{subfigure}[b]{0.45\textwidth}
        \includegraphics[width=\textwidth, angle=0]{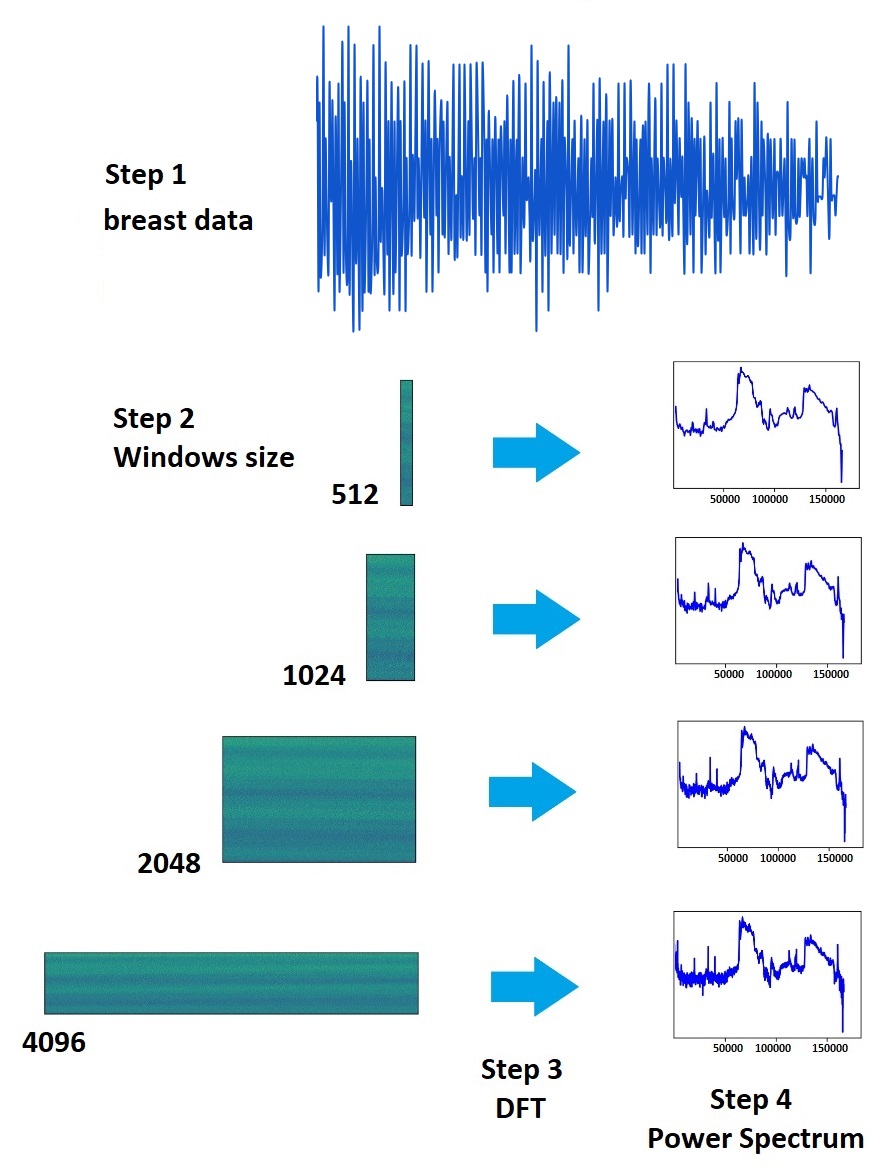}
        \caption{}
        \label{fig:normal_breast}
    \end{subfigure}
    \caption{Biological signals recorded with FEDSA from participants with (a) normal breast tissue, and (b) with breast abnormalities. In Step 1, data was captured from each sensor (D1 and D2) along with their analog difference (Dif). Step 2 involved organizing the data appropriately (see Figure~\ref{tratamientodedatos.PNG} for details). The Fourier transformation of the data was performed in Step 3, followed by the calculation of the power spectrum in Step 4, as outlined in the Materials and Methods section.}

    \label{fig:PSN_combined}
\end{figure*}

In this study, the FEDSA system was used to analyze breast tissue samples with normal and abnormal diagnoses; see Figure~\ref{fig:PSN_combined}. The power spectrum was obtained by different window sizes (512, 1024, 2048, and 4096), observing similar behavior. It was observed that decreasing the window size to 512 led to a reduction in minor noise. This reduction in noise can be attributed to an increase in the number of individual windows included in the calculation, denoted as $\xi$ in Eq.~(\ref{eq:S-matrix}). Although when $\xi$ increases, the noise decreases, it is important to note that the number of spectral points ($l$) decreases.

\begin{table*}[t]
\caption{Statistical significant results ($p < 0.05$) obtained from the analysis of the \textit{bands} $b_i^{(k)}$, for the channel Dif, assessed using either the t-student test or the non-parametric independent-samples Mann-Whitney U test.}
\label{table:results}
\centering
\begin{tabular}{ccccc}
\hline 
Window & Breast & Quadrant & p-value & Test \\ \hline \hline 
512 & Right & UIQ  &  $0.047 $ & t-student
\\
& Right & LOQ &  $0.042$ & t-student
\\
& Left & UIQ & $0.030 $ & Mann-Whitney U 
\\
& Left & LOQ &  $0.016$  & t-student
\\\hline
1024 & Left & UIQ & $0.030$ & Mann-Whitney U 
\\
& Left & LOQ & $0.021$  & Mann-Whitney U 
\\\hline
2048 & Left & UIQ & $0.039$ & Mann-Whitney U 
\\
& Left & LOQ & $0.021$ & Mann-Whitney U 
\\\hline
4096 & Left & UIQ &  $0.045$ & Mann-Whitney U 
\\
& Left & UOQ & $0.043$ & t-student
\\
& Left & LOQ & $0.021$ & Mann-Whitney U  \\\hline 
\end{tabular}
\end{table*}

The results of the statistical analysis for the \textit{bands} $b _i^{(k)}$, for the Dif signal, with p-values lower than 0.05 are presented in Table~\ref{table:results}. Among the data from the three signal sources (D1, D2, and Dif), D1 and D2 did not show significant differences. However, data from Dif, which represents analogical subtraction between signals measured by detectors D1 and D2, showed significant differences in a specific frequency \textit{band} $b_6^{(k)}$ (150-160 kHz) between samples classified as normal and abnormal. Significant differences were predominantly observed in the left breast within the upper inner quadrant (UIQ) and lower outer quadrant (LOQ), regardless of the number of windows used to calculate the power spectrum. Furthermore, significant differences were observed in the UIQ and LOQ quadrants for the right breast.

We used PCA analysis to derive five independent variables, denoted as $\mathfrak{b}_j^{(k)}$, which served as input for a logistic regression model given by Eq.~(\ref{eq:log_reg_model}). The transformation matrix elements $T_{ji}$ used in the PCA analysis are presented in Table~\ref{tab:PCA_Matrix}, and the VIF for the five independent variables was one, indicating that there is no multicollinearity between them. In this analysis and in the following, we included data from all quadrants.

\begin{table*}[th]
\centering
\caption{Linear transformation matrix elements $T_{ji}$ of Eq.~(\ref{eq:LinearTransformationPCA}) obtained from the PCA analysis. Each row corresponds to a PCA variable $\mathfrak{b}_j^{(k)}$, while each column corresponds to an original variable $b_i^{(k)}$.}
\label{tab:PCA_Matrix}
\begin{tabular}{c||cccccc}
\hline
 & $b_1^{(k)}$ & $b_2^{(k)}$ & $b_3^{(k)}$ & $b_4^{(k)}$ & $b_5^{(k)}$ & $b_6^{(k)}$ \\
\hline \hline
$\mathfrak{b}_1^{(k)}$ & -0.01015866 & -0.16355728 & -0.69739404 &  0.47083992 &  0.51035184 & -0.06810391 \\
$\mathfrak{b}_2^{(k)}$ &  0.01522481 &  0.00748132 & -0.06192288 & -0.75415790 &  0.63466554 &  0.15595930 \\
$\mathfrak{b}_3^{(k)}$ & -0.01453149 & -0.08667928 & -0.35824755 & -0.07405752 & -0.33395900 &  0.86424598 \\
$\mathfrak{b}_4^{(k)}$ &  0.15250132 &  0.81396142 & -0.44663829 & -0.17280750 & -0.21346087 & -0.19823324 \\
$\mathfrak{b}_5^{(k)}$ &  0.90060133 & -0.33748231 & -0.12147502 & -0.13245745 & -0.15262245 & -0.13938495 \\
\hline
\end{tabular}
\end{table*}

To assess the reliability of these variables, we analyzed the explained variance ratios (EVR) obtained from the principal component analysis (PCA). The EVR measures the proportion of variance captured by each variable, providing information on its significance. Our analysis revealed that the first two PCA variables, $\mathfrak{b}_1^{(k)}$ with an EVR of 0.8260 and $\mathfrak{b}_2^{(k)}$ with an EVR of 0.1672, represented the majority of the variance in the dataset. Specifically, $\mathfrak{b}_1^{(k)}$ explained approximately 82.6\% of the variance, while $\mathfrak{b}_2^{(k)}$ explained around 16.7\% of the variance. However, the remaining PCA variables, $\mathfrak{b}_3^{(k)}$ with an EVR of 0.0042, $\mathfrak{b}_4^{(k)}$ with an EVR of 0.0022, and $\mathfrak{b}_5^{(k)}$ with an EVR of 0.0004, contributed relatively smaller proportions to the total variance; these results underscore the critical role played by the first two PCA variables in the capture of essential information and patterns within the dataset, while the remaining variables make smaller contributions to the overall variance.

We estimate the sensitivity and specificity of FEDSA in differentiating between normal and abnormal breast tissue using logistic regression and ROC analysis. It is important to note that the ROC analysis provides an estimate rather than a conclusive result due to the number of data collected in this study. However, it serves as a suitable proof of concept for the research presented in this study. 

The coefficients ($\beta_n$) of the logistic regression model indicate the strength and direction of the relationship between each predictor variable and the target variable. The coefficient obtained for the constant term $\beta_0$ is -1.3928, suggesting a negative influence on the target variable. The coefficients for the PCA variables ($\mathfrak{b}_1^{(k)}$ to $\mathfrak{b}_5^{(k)}$) are $\beta_1=87.4380$, $\beta_2=-70.5654$, $\beta_3=-631.8344$, $\beta_4=475.0971$, and $\beta_5=1523.8956$. The significance of each coefficient is determined by the p-value, with values less than 0.05 indicating statistical significance. In this case, all predictor variables, including the constant term, have p-values less than 0.05, indicating their significance in the logistic regression model.

\begin{figure}[t]
    \centering
    \includegraphics[width=0.45\textwidth]{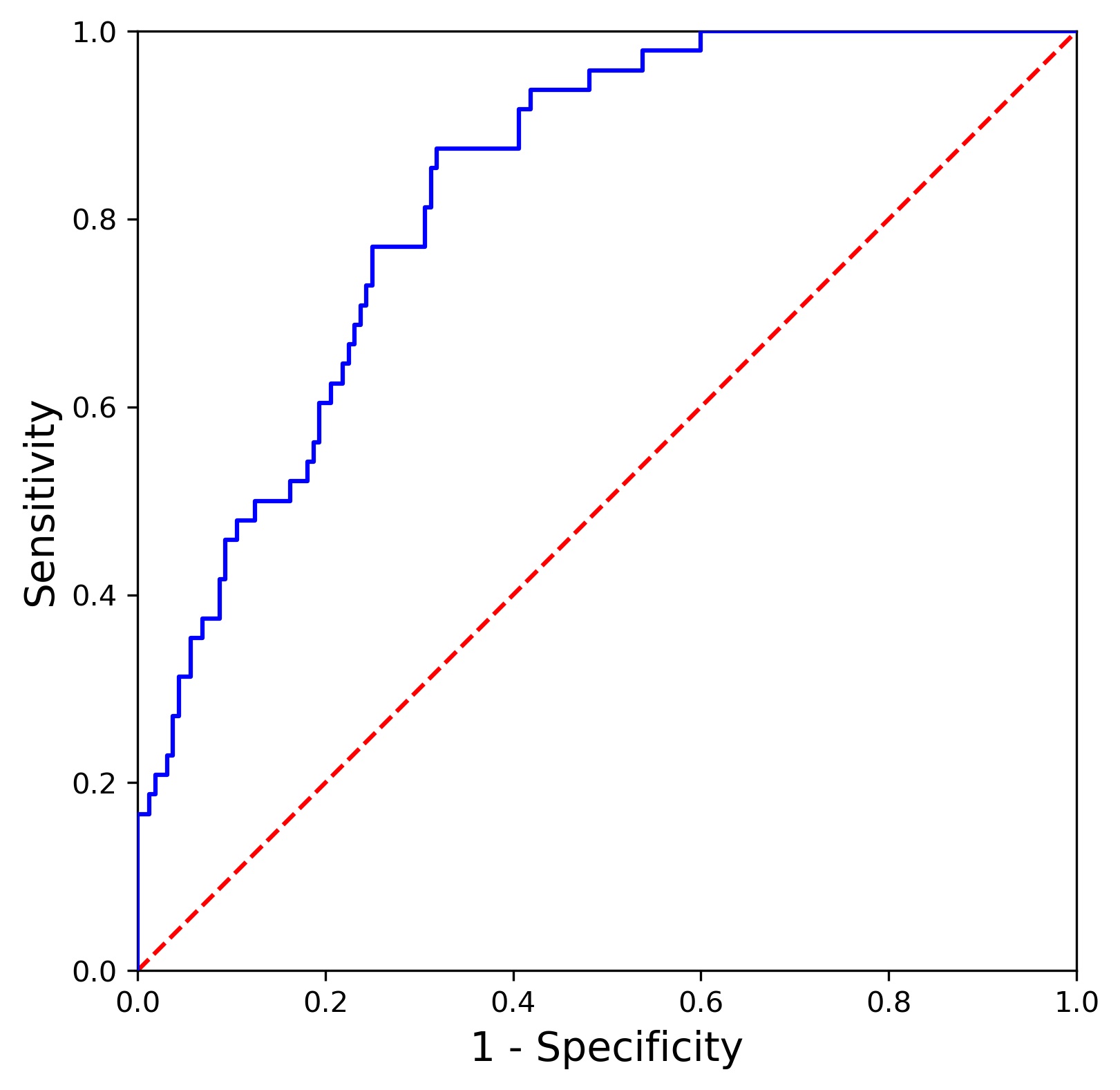}
    \caption{ROC curve for the logistic regression predicted variable $\mathfrak{p}_k$, as defined by Eq.~(\ref{eq:log_reg_model}), with a cutoff of 0.1905. The curve demonstrates a sensitivity of 0.88 and a 1-specificity of 0.32. The area under the curve (AUC) is calculated to be 0.83.}
    \label{fig:roc_curve}
\vspace{\baselineskip}
\end{figure}

The logistic regression model, described by Eq.(\ref{eq:log_reg_model}) and with the parameter values $\beta_n$ mentioned above, underwent a ROC analysis to assess its performance in distinguishing between normal and abnormal conditions of breast tissue (Figure~\ref{fig:roc_curve}). Sensitivity and specificity measures were used to assess the ability to correctly identify positive and negative cases, respectively, using the logistic regression model. The optimal cut-off point was determined using the Youden J statistic, maximizing the overall accuracy of the model in classifying breast tissue conditions. The cutoff value obtained at 0.1905 yielded a sensitivity of 87.5\%, indicating a high capacity to accurately identify abnormal breast tissue using the logistic model. The specificity obtained was 68.1\%, indicating its precision in classifying cases of normal breast tissue. Furthermore, the area under the curve (AUC) was calculated at 0.83, demonstrating the discriminatory power of the logistic regression model to distinguish between the two classes. The AUC value implies that the model performs well in distinguishing between normal and abnormal breast tissue conditions, with a higher likelihood of correctly classifying randomly selected instances from the abnormal class than from the normal class.\footnote{Processed experimental data and a simple Jupyter Notebook that show the data processing is available at \cite{DavidAM}.}

\section{Discussion}

Malignant transformation of breast tissue has traditionally been attributed to genetic \cite{danforth2016genomic,lunardi2022genetic} and epigenetic \cite{muhammad2022potential,joshi2022epigenetic,khan2022epigenetics} modifications. These alterations affect protein synthesis and behavior, leading to changes in biochemical and molecular signaling within tissue \cite{zhang2023molecular,deepak2020tumor}. Consequently, this affects the acquisition of nutrients, oxygen utilization, cell death, glucose metabolism, oncogene activation, tumor suppressor gene inactivation, apoptosis control, and cell cycle regulation \cite{lehmann2011immunohistochemical, gojis2010role,cerma2023targeting}. Such disruptions influence not only transformed cells, but also the cellular environment. Understanding and monitoring these changes is crucial due to the heterogeneity of breast cancer. The concept of the field cancerization effect, which encompasses all biological, physical, chemical, and morphological changes during tissue transformation, has been instrumental.

FEDSA was proposed as a proof-of-concept to measure the amplification of the field effect in breast tissue. The acquisition system and data analysis approach of FEDSA aim to capture changes in tissue dynamics, which, according to Brownian motion theory, contribute to the frequency power spectrum associated with the sizes of internal tissue components.

Two experiments validated this proof of concept. The first involved suspended nanoparticles grouped by size. Analysis of the power spectrum of these nanoparticles as a function of frequency, shown in Figure~\ref{SPP.png}, demonstrated that the decay curve of the power spectrum tends toward higher frequencies for smaller nanoparticles. This behavior is consistent with the findings obtained using the DLS technique \cite{hassan2015making}. Additionally, the standard deviation of the power spectrum for each group of nanoparticles was calculated, and the results from the calibration sample of polystyrene nanoparticles were within the standard deviation predicted by FEDSA. This confirmed FEDSA's ability to measure the contribution of nanoparticles to the power spectrum as a function of frequency, with different nanoparticle sizes resulting in distinct power spectrum patterns. This analogy between the movement of the cellular components within the tissue and the suspension of particles in water was based on the random movement of these elements within the tissue and the predominance of water in the tissues.

The second experiment involved the analysis of normal and abnormal breast tissue samples from 17 and 7 women, respectively. Analysis of the power spectrum as a function of frequency revealed significant differences between the five quadrants of breast tissue, as presented in Table~\ref{table:results}. The FEDSA illumination system used a $650 nm$ wavelength light beam, known to penetrate up to 5 cm into breast tissue \cite{hall2015guyton, tuchin2015tissue}. Consequently, the frequency bands that differentiated between normal and abnormal tissue states potentially contained valuable information derived from backscattered light that interacts with molecules in various cellular and extracellular structures. These structures include fat cells, collagen fibers, epithelial cells, microtubules, leukocytes, cell nuclei, elastin fibers, mitochondria, lysosomes, and collagen fibrils. The power spectrum analysis thus provided information on variations in tissue composition and organization associated with different tissue states, thereby enhancing our understanding of the biological and morphological changes that occur in breast tissue.

The integration of PCA analysis, logistic regression modeling, and ROC analysis significantly improved our understanding and predictive capabilities of the FEDSA technique. PCA analysis was crucial in reducing the dimensionality of the dataset by extracting independent variables that captured a substantial portion of the variance of the dataset. These variables were then used as input for the logistic regression model, which demonstrated remarkable discriminatory power in distinguishing between normal and abnormal breast tissue conditions. Subsequent ROC analysis provided a comprehensive evaluation of the model's performance by assessing its sensitivity and specificity in correctly classifying positive and negative cases. The determination of the optimal cutoff point, based on the Youden J statistic, ensured an optimal balance between sensitivity and specificity, maximizing the overall precision of the model in classifying breast tissue conditions. However, due to the limited sample size (abnormal breast tissue samples from seven women and normal breast tissue samples from 17), the ROC analysis results provide a preliminary indication of the concept's potential. More experiments with larger datasets are needed to obtain more precise cut-off values, sensitivity, and specificity of the FEDSA technique to discriminate between normal and abnormal breast tissue. Taking into account the sensitivity and specificity values obtained from the current analysis, we can state that the concept has been successfully tested, providing valuable information for future investigations and improvements in the application of FEDSA for the analysis of breast tissue.

\section{Conclusions}

This proof-of-concept study holds significant promise in implementing and further developing FEDSA as a valuable tool for breast tissue analysis. The technique offers several advantages, including its low cost, non-ionizing radiation nature, portability, and ability to differentiate between normal and abnormal tissue states within the study group of women. Furthermore, FEDSA presents a unique opportunity to explore and identify potential biomarkers to detect early changes in breast tissue, addressing the current lack of technologies capable of detecting these initial changes, often not palpable or visible by mammography. Interestingly, our data analysis revealed a quadrant-specific dependence, highlighting the need for further studies and statistical analyzes specific to the Colombian population to establish relationships between breast cancer incidence and the appearance of alterations in particular quadrants of breast tissue. By addressing these research gaps, FEDSA has the potential to contribute to the evaluation of breast tissue and to the early detection and improved treatment of breast cancer in our population.

\appendix


\printcredits

\section*{Supplementary information}
Processed experimental data and a simple Jupyter notebook showing the data processing are available at \cite{DavidAM}.

\section*{Acknowledgements}
The authors appreciate the support received for this work from the \textit{Universidad Industrial de Santander} under Grant VIE 2816. One of the authors, JFP, extends thanks for the support provided by a MinCiencias Doctoral Scholarship. Our sincere appreciation is also extended to the Leam spectroscopy laboratory for granting access to the dynamic light scattering equipment and for providing the necessary space for data collection at the \textit{Empresa Social del Estado, Hospital Universitario de Santander} (ESE-HUS).

%

\bibliographystyle{model1-num-names}

\bibliography{biblio}



\end{document}